\documentstyle[eqsecnum,prb,aps,twocolumn]{revtex} 
\begin{document}
\baselineskip 0.525cm
\title{\bf Magnetic behavior of Eu$_2$CuSi$_3$\\ Large negative
magnetoresistance above Curie temperature}

\author{Subham Majumdar, R. Mallik and E.V. Sampathkumaran$^*$}

\address{Tata Institute of Fundamental Research, Colaba, Mumbai-400005,
INDIA}

\author{Kirsten Rupprecht and G. Wortmann }   % Use this and the next line only if there is a second

\address{Fachbereich 6-Physik, Universitaet Paderborn, D-33095, GERMANY}
% address. (Remove the left % marks) %
\maketitle

\begin{abstract}
\noindent We report here the results of magnetic susceptibility,
electrical-resistivity, magnetoresistance (MR),  heat-capacity and
$^{151}$Eu M\"ossbauer effect measurements on the compound,
Eu$_2$CuSi$_3$,  crystallizing in an AlB$_2$-derived hexagonal structure.
The results establish that Eu ions are divalent, undergoing long-range
ferromagnetic-ordering below (T$_C$=) 37 K. An interesting observation is
that the sign of  MR is negative even at temperatures close to 3T$_C$,
with  increasing magnitude with decreasing temperature exhibiting a peak
at T$_C$. This observation, being made for a Cu containing  magnetic
rare-earth compound for the first time, is of relevance to the field of
collosal magnetoresistance.
\end{abstract}
\vskip 0.2cm

%\vskip 0.4cm
\noindent Large negative magnetoresistance (LNMR) in perovskite-based manganites
[colossal magnetoresistance (CMR) systems] peaking in the vicinity of
Curie temperature (T$_C$) is one of the most important observations  in
modern condensed matter physics.\cite{1}  In this regard, the observation
of negative magnetoresistance, the magnitude of which increases with
decreasing temperature peaking near the magnetic ordering temperature
temperatures (T$_o$) in the indirect-exchange controlled  (not only in
ferromagnetic, but also in antiferromagnetic) magnetic systems like those
of Gd, Tb and Dy based alloys (by us\cite{2,3,4,5,6}) calls for mechanisms
other than double-exchange and Jahn-Teller effects to explain such
features. We have proposed the need to consider the possible role of
magnetic polaronic effects even in metallic alloys. The importance of
such an observation and hypothesis is apparent also from similar recent
reports, both theoretical and experimental, from other groups as
well.\cite{7,8,9,10,11} Apparently, it is not always necessary to observe
such features  in every Gd alloy and in fact many such alloys show
normally expected negligibly small positive magnetoresistance (MR) at all
temperatures above T$_o$.\cite{6} However, it was noted that such features,
if observed, are restricted to Pd, Pt, Co, Ni or Mn containing compounds,
but not to Cu, Ag and Au containing ones, which raise a question whether
such anomalies result from the intrinsic tendency of the d bands of the
former class of elements  to get easily polarised by the 4f local moment.
In this article, among other magnetic properties, we emphasize on the
observation of LNMR in a (Eu-based) Cu containing alloy for the first time
over a wide range of temperature above Curie temperature (T$_C$); Cu being
non-magnetic, this result endorses our view that  LNMR is a magnetic
precursor effect of 4f-ion long-range magnetic ordering.

The  investigation of  Eu$_2$CuSi$_3$, crystallizing in an AlB$_2$-type
hexagonal structure has  been  undertaken   considering current interest
in synthesizing and investigating  magnetic properties of ternary
rare-earth/actinide compounds with the atomic ratios of 2:1:3,
particularly those adopting variants of either the tetragonal ThSi$_2$ or
hexagonal AlB$_2$ structure
types.\cite{4,5,12,13,14,15,16,17,18,19,20,21,22,23} It is to be noted
that the reports on Eu compounds of this type are generally rare.
Recently, we reported the synthesis and interesting magnetic behavior of
the compound, Eu$_2$PdSi$_3$ [Ref. 19]. The scarcity of the reports on
such Eu compounds is presumably due to  the difficulties in controlling
the Eu stoichiometry during sample preparation. We have carried out
electrical resistivity ($\rho$), MR, magnetization (M) and magnetic
susceptibility ($\chi$), heat-capacity (C) and $^{151}$Eu M\"ossbauer
effect (ME) studies  on Eu$_2$CuSi$_3$, in order to understand the
magnetic behavior of this alloy, the results of which are reported here.

The sample was prepared by induction melting stoichiometric amounts of
constituent elements in an inert atmosphere. The ingot was melted four
times and the loss due to evaporation of Eu after first melting was
compensated by adding corresponding amounts of Eu. We noticed negligible
loss of Eu in subsequent meltings. The ingot was homogenised in an
evacuated, sealed quartz tube at 800$^o$C. X-ray diffraction pattern,
obtained by employing Cu K$_\alpha$ radiation, established that the
present compound  crystallizes in an AlB$_2$-derived hexagonal structure.
Since we do not see any superstructure lines in the x-ray diffraction
pattern, we presently believe that there is an intrinsic disorder between
Cu and Si sites, unlike Eu$_2$PdSi$_3$ (Ref. 19),  and the lattice
constants obtained are a= 4.095\AA\hskip2mm  and c=4.488\AA.   No
additional lines attributable to any other phase within the detection
limit of x-ray diffraction could be seen.

The $\rho$ measurements (2-300 K) were performed by a conventional
four-probe method employing silver paint for making electrical contacts.
The sample is found to be very porous and tends to become powder with
ageing and hence there are difficulties at arriving absolute values of
$\rho$. The $\chi$ measurements (2-300 K) were performed employing a
commercial SQUID magnetometer in the presence of  several magnetic fields.
The C data (2-70 K) were obtained by a semi-adiabatic heat-pulse method.
The (longitudinal mode) MR data were obtained in the presence of a
magnetic field (H) of 30 kOe in the temperature range 4.2-100 K and also
as a function of H at selected temperatures (4.2, 25 and 50 K). $^{151}$Eu
ME measurements  at selected temperatures (4.2 - 300 K) were performed
employing $^{151}$SmF$_3$ source (21.6 keV transition) in the transmission
geometry.

The results of $\rho$ (normalised to the value at 300 K), inverse $\chi$
(measured in the presence of 2 kOe) and C measurements are shown in Fig.
1, only below 100 K; the data at higher temperatures are not shown as
there are no interesting features to be highlighted. It is clear that
there is a sudden drop in $\rho$ at  37(1) K as the temperature is
lowered. This drop arises from the onset of magnetic ordering as evidenced
below.  There is a distinct anomaly even in the temperature dependent C
data around 37 K, which originates from magnetic ordering.  The $\chi$ is
found to exhibit Curie-Weiss behaviour in the temperature range 40-300 K
and the effective moment obtained from this linear region is found to be
7.8 $\mu$$_B$/Eu, which is very close to that expected for divalent Eu
ions, thereby confirming that all the Eu ions are divalent in this
compound. This also suggests that there is no magnetic moment on Cu. The
value of the paramagnetic Curie-temperature ($\theta$$_p$) is found to be
38 K; inverse $\chi$ tends to saturate below the same temperature. These
results establish that Eu ions undergo a ferromagnetic-type of magnetic
ordering at 37(1) K;  this temperature is practically the same as
$\theta$$_p$, indicating absence of competition from antiferromagnetic
interaction, a situation different from that observed for Eu$_2$PdSi$_3$
(Ref. 19).

There are also qualitative changes in the low temperature susceptibility
behavior measured at different fields, as seen in Fig. 2. In addition to
the sharp rise of $\chi$ around 40 K due to the onset of ferromagnetic
ordering, there is a peak at about 5 K followed by a drop at lower
temperatures for the data recorded  in the presence of 100 or 1000 Oe, a
feature absent (but showing a weak upturn) for the application of a higher
field (say, 2 kOe). This
might imply the presence of another magnetic transition around 5 K, which
is modified with increasing magnetic field. This may be corroborated to
the observation of a very weak peak in C around 5 K (visible if the plot
of the low temperature data is drawn in an expanded scale). Presumably,
the two magnetic transitions arise from two types of Eu ions with
different chemical environment, which may be intrinsic to these 2-1-3
class of alloys;\cite{19} possibly the degree of Cu-Si disorder is
actually small. The evidence for two types of Eu ions can be found even in
the M\"ossbauer data and the broadening of the spectra of the minority
site due to the transferred hyperfine field from the majority site
(discussed below) establishes that minority Eu is not extrinsic to the
sample. We also note that the zero-field and field-cooled data diverge at
T$_C$ as the temperature is lowered, presumably due to anisotropy of the
material; this divergence can not arise from spin-glass phenomenon,
considering that the features in C and $\rho$ are sharp at the magnetic
transition.  We may also add that there is a hysteretic behavior of the
isothermal M with a small coercive field (300 Oe) at 2 K, but with a still
smaller value of coercive field at 10 K,  if measured as a function of H
(Fig. 3), typical of soft ferromagnets. It is obvious from Fig. 3 that the
isothermal M does not saturate even at high fields and the value, say  at
2 K for H= 50 kOe is far below the full value of 14$\mu$$_B$/formula unit;
which may suggest that the ferromagnetism could be of a canted-type.

We have also performed $^{151}$Eu M\"ossbauer effect measurements as a
function of temperature in order to get a microscopic picture of the
magnetism.  It may be remarked that the $^{151}$Eu ME studies at 300 K for
the composition, EuCu$_{0.5}$Si$_{1.5}$, was reported in Ref. 24 several
years ago, as a continuation of substitutional studies in EuSi$_2$, but to
our knowledge there has been  no further study on this alloy. Typical
spectra obtained at various temperatures are shown in Fig. 4, reflecting
the magnetic behavior below 38 K. The spectrum at 38 K indicates the
absence of magnetic order at this temperature, with the dominant feature
at -10.5 mm/s characteristic of divalent Eu ions. A weak feature around
zero velocity ( at  - 0.2 mm/s) with about 3\% fractional intensity (as
derived from the low temperature spectra) arises from trivalent Eu ions,
presumably produced by surface oxidation when powdering the sample for the
preparation of the absorber. The magnetic hyperfine split spectra  below
38 K could not be consistently fitted with only one site; the assumption
of two sites with the same intensity ratio of 3:1 as in Eu$_2$PdSi$_3$
(Ref. 19), however, resulted in a better description of the observed data.
The isomer shifts of the two sites are almost identical (-10.4 and -10.7
mm/s), in contrast to distinctly different values (-10.0 and -8.6 mm/s)
observed in Eu$_2$PdSi$_3$. The hyperfine fields, B$_{eff}$, of the two
sites exhibit , similar to the  Eu$_2$PdSi$_3$ case, a drastically
different temperature dependence, as displayed in Fig. 5. At 4.2 K, the
values of B$_{eff}$,  -290 and -315 kOe, are rather close for the majority
and minority sites, unlike the situation in  Eu$_2$PdSi$_3$ (in which case
the corresponding values are -408 and -255 kOe respectively). Since the
unit-cell volumes of these Cu and Pd containing alloys are almost
identical, the differences in isomer shifts and hyperfine fields between
these two compounds may be attributed to different  transition metal ion
surrounding for Eu. Obviously, the influence of Pd is stronger in bringing
about  a local modification of the conduction electron characteristics
around the two Eu sites. In addition, Pd and Cu environments also modify
the nature of magnetic ordering; while in the Pd case there is an evidence
for antiferromagnetic coupling, the magnetic ordering is of a
ferromagnetic type in the Cu case. It should also be noted that the plot
of B$_{eff}$ versus temperature does not follow the magnetization curve
expected for a S=7/2 spin system, which may  be indicative of some degree
of crystallographic disorder. There is a transfer of  hyperfine field to
the minority  Eu site below 38 K as indicated by the broadening of the
spectral features of this site, however, with the minority site ordering
magnetically only at lower temperatures. Around 6 K, similar to a feature
in the $\chi$ data, there is a noticeable change in the  B$_{eff}$(T)
curves for both the sites, indicating a rearrangement of the spin
direction, possibly resulting in a canted ferromagnetic alignment. From
the absence of a significant reduction in the value of B$_{eff}$ below
that due to core-polarization contribution, following the arguements given
in our earlier article,\cite{19} we conclude that the antiferromagnetic
coupling is however negligible. This is consistent with the observation
that, besides the sign, the magnitude of  $\theta$$_p$ is the same as
T$_C$.

We now present an observation of  importance to the field of CMR.  An
application of H (say, 30 kOe) depresses the value of the electrical
resistance, resulting in a negative MR even around 100 K, which is far
above T$_C$ (Fig. 6). The magnitude of MR, defined as $\Delta\rho/\rho$=
[$\rho$(H)-$\rho$(0)]/$\rho$(0), grows with decreasing temperature peaking
at T$_C$, as shown in Fig. 6, similar to the behavior observed in by now
well-known  perovskites-based La manganites (CMR systems). Needless to
mention that the negative MR in the ferromagnetically ordered state is
usually expected.  The data were also collected (Fig. 6,  bottom) as a
function of H at selected temperatures both above and below T$_C$ in order
to correlate the values with those obtained from the temperature dependent
measurements. It is clear that the value is as large as about -9\% at 70
kOe even at 50 K. We also remark that there is a sign crossover around 120
K, beyond which one sees a positive MR; for instance, in a field of 30 kOe
at 200 K, the value is about 2\%, which reduces to negligibly small (but
positive) values at 300 K (not shown in the figure).

To conclude, the compound, Eu$_2$CuSi$_3$, crystallizing in a
AlB$_2$-derived hexagonal structure, exhibits long-range ferromagnetic
ordering at 37 K. The temperature dependent magnetoresistance behavior is
qualitatively similar to that observed in CMR systems, though
double-exchange or Jahn-Teller mechanisms are not operative in such Eu
systems. Similar temperature dependent behavior in the paramagnetic state
has been made by us in the past in some Gd, Tb, Dy based
alloys.\cite{2,3,4,5,6} The uniqueness of the present result is that such
an observation is made for the first time in a Eu-based / Cu-based  alloy.
The Cu ions  do not carry any magnetic moment and therefore the present
results establish that the observed magnetoresistance anomalies in the
vicinity of magnetic transition temperature is solely related to magnetic
precursor effect due to 4f magnetism. We suggest that the alloys
exhibiting such features should be viewed together with LNMR systems to
understand this phenomenon better. The results also give a clue to
identify potential candidates  exhibiting large MR at room temperature for
applications in the sense that one can search for magnetic precursor
effects in compounds undergoing magnetic ordering at temperatures rather
close to (but below) 300 K.

The authors thank R. P\"ottgen for a discussion. 

\vskip 0.5cm

\rightline{$^*$Electronic address: sampath@tifr.res.in}

\begin{figure}
% if it is an e-mail submission. (Good photo or India ink drawing.)
\caption{Inverse susceptibility, electrical resistivity normalised to the
value at 300 K ($\rho(T)/\rho(300 K)$), and heat-capacity in the
temperature range 2 - 100 K for Eu$_2$CuSi$_3$.}

\end{figure}

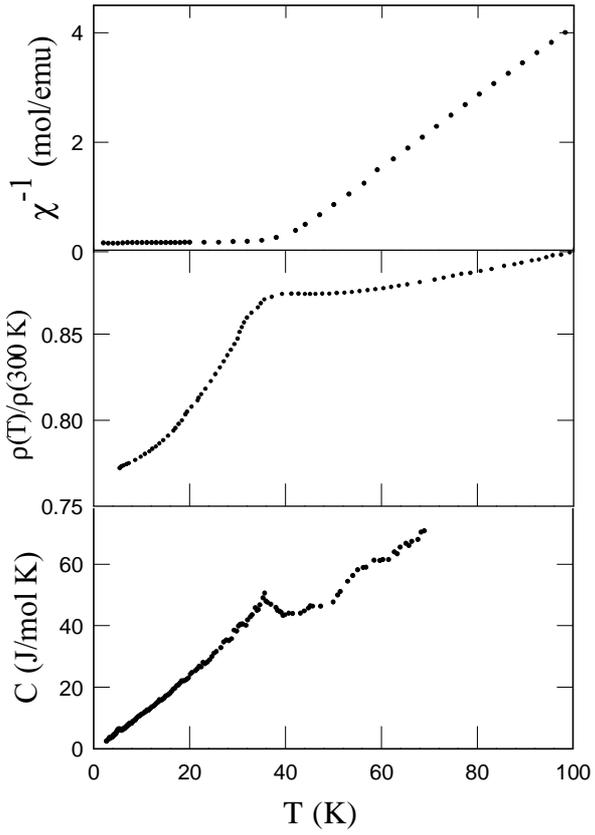
\begin{figure}

\caption{Magnetic susceptibility below 70 K measured in the presence of
various fields for Eu$_2$CuSi$_3$. All the data were recorded in the
zero-field cooled (ZFC) state of the specimen. For H= 100 Oe only, the
data in field-cooled (FC) state are also shown. Continuous lines are drawn
through the data points for H= 100 Oe.}

\end{figure}

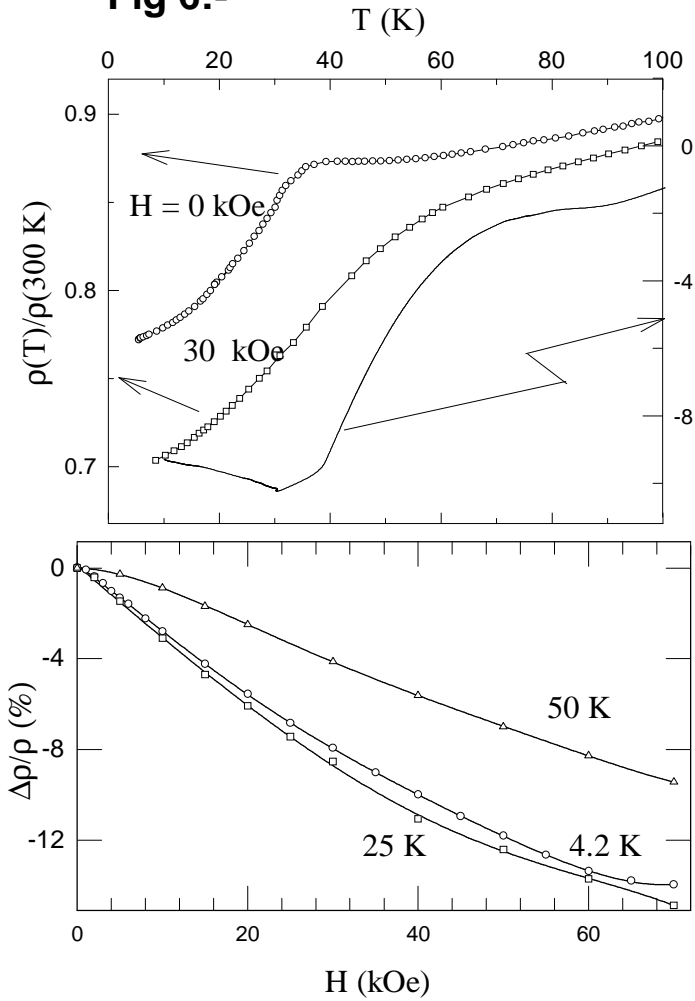
\begin{figure}

\caption{Magnetic hysteresis curves at 2 and 10 K for Eu$_2$CuSi$_3$. The
low field data is shown in an expanded form in the insets.The lines drawn
through the data points are guides to the eyes.}

\end{figure}

\begin{figure}

\caption{$^{151}$Eu M\"ossbauer spectra at various temperatures for
Eu$_2$CuSi$_3$. The continuous lines through the data points represent
least-square fits of the data. The subspectra for T $<$ 38 K are also
shown.}

\end{figure}

\begin{figure}

\caption{The magnitude of the  magnetic hyperfine fields plotted as a
function of temperature, obtained from the data shown in Fig. 4 for
divalent Eu ions with two different crystallographic environments.  Filled
circles are for majority Eu ions and the open circles are for minority Eu
ions. The lines through the data points are guides to the eyes. The sign
of B$_{eff}$ is negative.}

\end{figure}

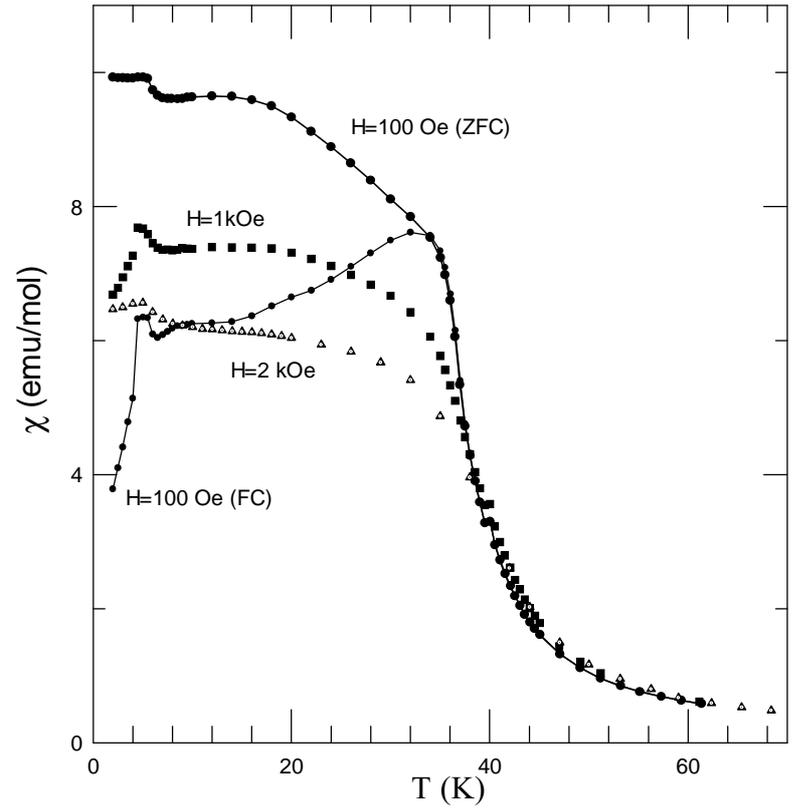
\begin{figure}

\caption{Electrical resistivity normalised to 300 K value as a function of
temperature for Eu$_2$CuSi$_3$ in zero field as well as in 30 kOe. The
magnetoresistance values derived from this data are also shown in this
figure.  The magnetoresistance as a function of field at selected
temperatures is shown in the lowest part of the figure.The lines drawn
through the data points serve as guides to the eyes.}

\end{figure}

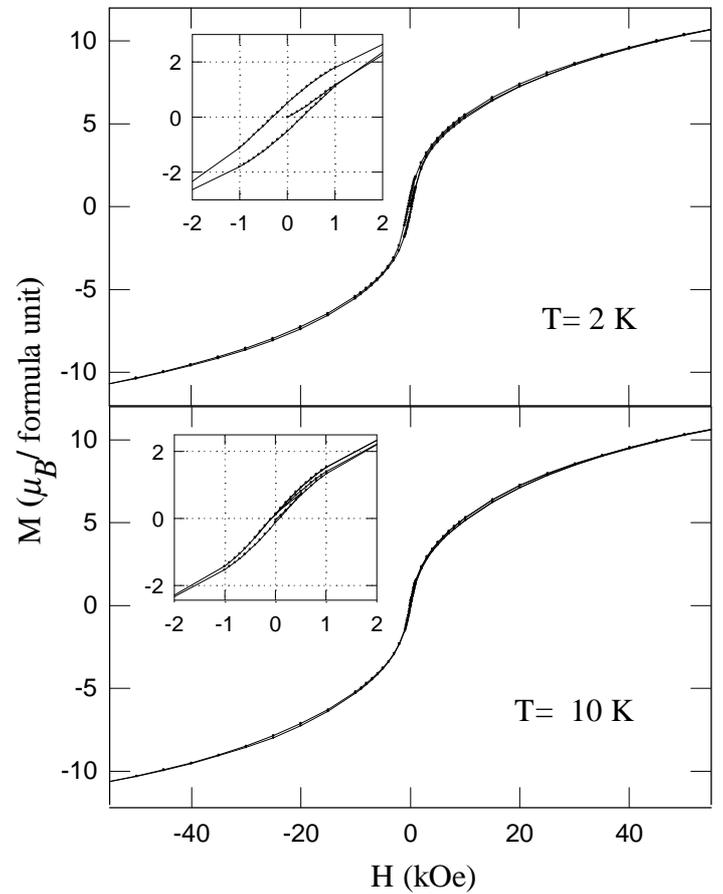
\end{document}